\documentclass{phb-proc4-auth}

\usepackage{graphicx}
\usepackage{amssymb}

\begin{document}
\begin{frontmatter}


\journal{SCES '04}


\title{Effect of hydrostatic pressure on the ambient pressure superconductor
CePt$_3$Si}


\author[DD]{M. Nicklas\corauthref{1}}
\author[DD]{G. Sparn}
\author[WI]{R. Lackner}
\author[WI]{E. Bauer}
\author[DD]{F. Steglich}


\address[DD]{Max Planck Institute for Chemical Physics of Solids,
N\"{o}thnitzer Str. 40, 01187 Dresden, Germany.}
\address[WI]{Institute of Solid State Physics, Vienna University of Technology,
Wiedner Hauptstrasse 8 - 10, A-1040 Wien, Austria.}

\corauth[1]{Corresponding Author: Max Planck Institute for Chemical Physics of
Solids, N\"{o}thnitzer Str. 40, 01187 Dresden, Germany.  Phone: +49 (0)351
4646 3127,    Fax: +49 (0)351 4646 3119, Email: nicklas@cpfs.mpg.de}


\begin{abstract}
We studied the evolution of superconductivity (sc) and antiferromagnetism (afm) in the
heavy fermion compound CePt$_3$Si with hydrostatic pressure. We present a
pressure-temperature phase diagram established by electrical transport measurements.
Pressure shifts the superconducting transition temperature, $T_c$, to lower
temperatures. Antiferromagnetism is suppressed  at a critical pressure $P_c\approx0.5$
GPa.
\end{abstract}


\begin{keyword}
CePt$_3$Si \sep superconductivity \sep antiferromagnetism \sep hydrostatic
pressure
\end{keyword}


\end{frontmatter}


Superconductivity (sc) is one of the most striking effects in solid state
physics. In a conventional superconductor Cooper pairing is mediated by
phonons. In general, magnetism destroys superconductivity. In heavy fermion
systems, however, sc exists in close proximity to magnetism, promoting the
suspicion that the sc is mediated by magnetic excitations. Since the discovery
of sc in the heavy fermion compound CeCu$_2$Si$_2$ at atmospheric pressure
\cite{Steglich79}, only a few Ce-based systems were found which also exhibit sc
at atmospheric pressure, like CeMIn$_5$ (M=Co, Ir) \cite{Thompson03}. Most
superconducting pure Ce-based systems show sc only under applied pressure
sufficient to suppress long range magnetic order, like CeIn$_3$
\cite{Mathur98} or CeRh$_2$Si$_2$ \cite{Movshovich96}. CeIn$_3$ displays a
typical temperature-pressure phase diagram for these compounds;
antiferromagnetism (afm) is suppressed to zero temperature with pressure and sc
develops right in the vicinity where afm disappears \cite{Mathur98}. Very
recently another material, namely CePt$_3$Si, was found showing magnetic order
and sc at atmospheric pressure \cite{Bauer04}. In contrast to the systems
mentioned before, the crystal structure of CePt$_3$Si is non-centrosymmetric,
which is believed to allow for novel superconducting order parameter states.
In this work we investigate the pressure dependence of the superconducting
transition temperature, $T_c$, and of the N{\'e}el temperature, $T_N$, by
electrical resistivity, $\rho$, measurements to study the interplay of
magnetism and sc in CePt$_3$Si.

Polycrystalline $\rm CePt_3Si$ was prepared by high frequency melting,
followed by a heat treatment at 870$^{\circ}$C for 10 days. The phase purity
was checked by x-ray diffraction and electron microprobe measurements. $\rm
CePt_3Si$ crystallizes in a tetragonal structure with no center of inversion
symmetry. At ambient pressure CePt$_3$Si orders antiferromagnetically at
$T_N=2.2$ K and sc develops out of the antiferromagnetic state bellow
$T_c=0.75$ K \cite{Bauer04}. At higher temperatures, the resistivity shows two
pronounced curvatures at 75 K and 15 K \cite{Bauer04}. These features seem to
be related to crystal electric field effects in the presence of Kondo-type
interactions.

The sample was mounted in a clamp-type pressure cell with a 1:1 mixture of
n-pentane and 2-methyl-butane as pressure medium. The measured pressure shift
of the superconducting transition temperature of tin served as pressure gauge.
A standard 4-point Lock-In technique was used to measure the electrical
resistance.

The antiferromagnetic transition at 2.2 K leads to a change in slope of the resistivity
data, shown in figure 1. The well defined maximum in ${\rm d}\rho/{\rm d}T$ at $T_N$ is
used to follow the pressure dependence of $T_N$. Up to $P=0.36$ GPa $T_N$ changes little
with pressure, from $T_N=2.2$ K at ambient pressure to $T_N=1.9$ K at $P=0.36$ GPa. The
data at $P=0.61$ GPa show no indication of a magnetic transition anymore. A linear
extrapolation of $T_N\rightarrow0$ using the initial slope of $T_N(P)$, ${\rm
d}T_N(P)/{\rm d}P=(-0.9\pm0.2)$ K/GPa, leads to a critical pressure of $P_c\approx2.4$
GPa. Instead, afm is suppressed at much lower pressure, $P_c\approx0.5$ GPa, i.e., $T_N$
vanishes non-monotonically in a first order-like transition.

\begin{figure}[t]
\centering
\includegraphics[angle=0,width=60mm,clip]{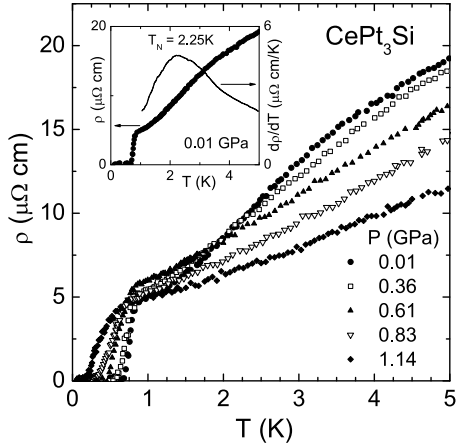}
\caption{Resistivity of CePt$_3$Si as a function of temperature for different
pressures. $T_c$ decreases with increasing pressure accompanied by a broadening
of the transition. Inset: resistivity and its derivative for the initial
pressure $P=0.01$ GPa. The maximum of the derivative indicates $T_N$.}
\end{figure}

In the paramagnetic phase at temperatures right above $T_N$ at atmospheric
pressure, the resistivity decreases with increasing pressure, which can be
attributed to an increase of the Kondo temperature $T_K$. At temperatures
bellow $T_N$ at ambient pressure, but above $T_c$, the behavior of
$\rho(P,T{\rm = constant})$ is more complicated. In this temperature range, an
increased scattering due to the shift of the magnetically ordered phase to
lower temperatures and the decrease due to the shift of $T_K$ to higher
temperatures compete. First, with suppressing $T_N$ the resistivity above
$T_c$ is increasing, but for pressures above $P=0.61$ GPa, $\rho(P,T{\rm =
constant})$ is decreasing again showing the same behavior like at higher
temperatures. This supports our conclusion that magnetism is vanishing in the
vicinity of $P_c\approx0.5$ GPa.

\begin{figure}[t]
\centering
\includegraphics[angle=0,width=65mm,clip]{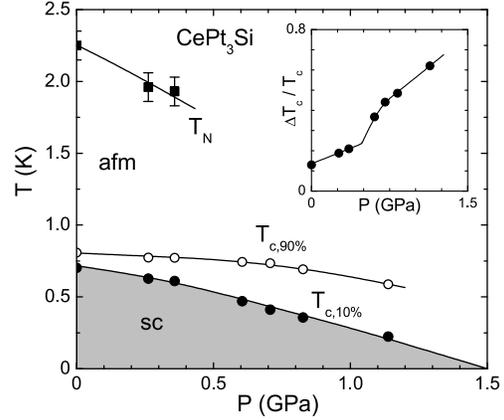}
\caption{Onset of the superconducting transition, $T_{c,90\%}$, defined by the
temperature where $\rho$ is 90\% of the normal state resistivity (open
circles), $T_{c,10\%}$ (closed circles) and N{\'e}el temperature $T_N$ (solid
squares). Inset: relative width of the superconducting transition $\Delta
T_c/T_c$. The lines are to guide the eye.}
\end{figure}

Superconductivity is persisting in a much broader pressure range than
antiferromagnetism in CePt$_3$Si. The initial slope ${\rm d}T_{c}/{\rm d}P= -0.18 \pm
0.03$ K/GPa of $T_c(P)$ is quite small. A smooth extrapolation of $T_c$ to higher
pressures gives $P^*\approx1.5$ GPa for the suppression of sc, as shown in figure 2.
Corroborated with results from doping studies, where Si is replaced isoelectronically by
Ge, corresponding to a negative chemical pressure \cite{Bauer05}, this indicates that
CePt$_3$Si is close to its maximum $T_c$ already at ambient pressure. This is different
from what is observed experimentally in, e.g., CeCu$_2$Si$_2$ \cite{Yuan03} or what is
expected from models of spin-fluctuation mediated sc, where $T_c$ becomes a maximum (or
minimum) at $P_c$. Even though no discontinuity can be resolved at $P_c$ a kink in the
relative width of the transition $\Delta T_c/T_c$ is observed at about $P_c$ (inset of
fig. 2). Since the width of the superconducting transition of tin shows no broadening,
this kink seems to be intrinsic to CePt$_3$Si and not caused by pressure
inhomogeneities.

We showed that with increasing pressure afm in CePt$_3$Si is suppressed
non-monotonically in a first order-like transition at $P_c\approx0.5$ GPa.
Superconductivity is not very sensitive to pressure and is persisting in a
broad pressure range up to $P^*=1.5$ GPa. No discontinuity of $T_c(P)$ is
observed at $P_c$. The maximum in $T_c(P)$ seems to be not related to $P_c$.
However, the superconducting transition width $\Delta T_c/T_c$ shows a
kink-like feature close to $P_c$.


\end{document}